\documentclass[iop]{emulateapj}

\def\spose#1{\hbox to 0pt{#1\hss}}
\def\simlt{\mathrel{\spose{\lower 3pt\hbox{$\mathchar"218$}}
     \raise 2.0pt\hbox{$\mathchar"13C$}}}
\def\simgt{\mathrel{\spose{\lower 3pt\hbox{$\mathchar"218$}}
     \raise 2.0pt\hbox{$\mathchar"13E$}}}

\providecommand\scription[2]{\scriptsize#1$\;${\scriptsize\uppercase\expandafter{\romannumeral #2}}\relax}%
\providecommand{\qso}{\ensuremath{\mbox{J\,0943+0531}}}
\providecommand{\qsofull}{\ensuremath{\mbox{SDSS~J\,094331.61+053131.4}}}
\providecommand{\galid}{\ensuremath{\mbox{227\_19}}}
\providecommand{\SFRnum}{\ensuremath{0.33\pm0.04}}
\providecommand{\HI}{\ensuremath{\mbox{\ion{H}{1}}}}
\providecommand{\scHI}{\ensuremath{\mbox{\scription{H}{1}}}}
\providecommand{\NHI}{\ensuremath{N_{\scHI}}}
\providecommand{\logNHI}{\ensuremath{\mbox{log\,}N_{\scHI}}}
\providecommand{\Lya}{\ensuremath{\mbox{Ly\,}\alpha}}
\providecommand{\Lyb}{\ensuremath{\mbox{Ly\,}\beta}}
\providecommand{\Lyc}{\ensuremath{\mbox{Ly\,}\gamma}}
\providecommand{\CII}{\ensuremath{\mbox{\ion{C}{2}}}}
\providecommand{\CIII}{\ensuremath{\mbox{\ion{C}{3}}}}
\providecommand{\scCIII}{\ensuremath{\mbox{\scription{C}{3}}}}
\providecommand{\NCIII}{\ensuremath{N_{\scCIII}}}
\providecommand{\logNCIII}{\ensuremath{\mbox{log\,}N_{\scCIII}}}
\providecommand{\SiII}{\ensuremath{\mbox{\ion{Si}{2}}}}

\providecommand{\OII}{\ensuremath{\mbox{\ion{O}{2}}}}
\providecommand{\OIII}{\ensuremath{\mbox{\ion{O}{3}}}}
\providecommand{\OV}{\ensuremath{\mbox{\ion{O}{5}}}}
\providecommand{\OVI}{\ensuremath{\mbox{\ion{O}{6}}}}
\providecommand{\SiIII}{\ensuremath{\mbox{\ion{Si}{3}}}}
\providecommand{\scOVI}{\ensuremath{\mbox{\scription{O}{6}}}}
\providecommand{\NOVI}{\ensuremath{N_{\scOVI}}}
\providecommand{\logNOVI}{\ensuremath{\mbox{log\,}N_{\scOVI}}}
\providecommand{\OVII}{\ensuremath{\mbox{\ion{O}{7}}}}
\providecommand{\MgII}{\ensuremath{\mbox{\ion{Mg}{2}}}}

\providecommand{\CIV}{\ensuremath{\mbox{\ion{C}{4}}}}
\providecommand{\NV}{\ensuremath{\mbox{\ion{N}{5}}}}

\providecommand{\Ha}{\ensuremath{\mbox{H}\alpha}}
\providecommand{\Hb}{\ensuremath{\mbox{H}\beta}}
\providecommand{\NII}{\ensuremath{\mbox{\ion{N}{2}}}}

\providecommand{\eV}{\,\ensuremath{\mbox{eV}}}
\providecommand{\Zsun}{\,\ensuremath{Z_{\odot}}}
\providecommand{\Myr}{\,\ensuremath{\mbox{Myr}}}

\providecommand{\ergs}{\,\ensuremath{\mbox{erg}\,\mbox{s}^{-1}}}
\providecommand{\arcsec}{\mbox{$^{\prime\prime}$}}%
\providecommand{\kmsMpc}{\,\ensuremath{\kms\,\Mpc^{-1}}}
\providecommand{\kpc}{\,\ensuremath{\mbox{kpc}}}
\providecommand{\Mpc}{\,\ensuremath{\mbox{Mpc}}}

\providecommand{\cmsq}{\,\ensuremath{\mbox{cm}^{-2}}}
\providecommand{\A}{\,\ensuremath{\mbox{\AA}}}
\providecommand{\mA}{\,\ensuremath{\mbox{m\AA}}}

\providecommand{\Msunyr}{\,\ensuremath{\mbox{M}_{\odot}\,yr^{-1}}}
\providecommand{\kms}{\,\ensuremath{\rm{km\,s}^{-1}}}
\providecommand{\dv}{\ensuremath{\Delta\,v}}
\providecommand{\zabs}{\ensuremath{z_{\mbox{abs}}}}
\providecommand{\zqso}{\ensuremath{z_{\mbox{qso}}}}

\providecommand{\zgal}{\ensuremath{z_{\mbox{gal}}}}
\providecommand{\Ls}{\ensuremath{\mbox{L}^{*}}}

\providecommand{\vLSR}{\ensuremath{v_{\mbox{\tiny LSR}}}}
\providecommand{\K}{\,\ensuremath{\mbox{K}}}
\providecommand{\logT}{\ensuremath{\mbox{log}\,T}}
\providecommand{\logU}{\ensuremath{\mbox{log}\,U}}
\providecommand{\Omegam}{\ensuremath{\Omega_{m}}}
\providecommand{\Omegal}{\ensuremath{\Omega_{\Lambda}}}

\providecommand{\Hn}{\ensuremath{\mbox{H}_{0}}}
\providecommand{\etal}{\ensuremath{\mbox{et~al.}}}
\providecommand{\dex}{\,\ensuremath{\mbox{dex}}}
\providecommand{\BV}{\ensuremath{B\!-\!V}}
\providecommand{\EBV}{\mbox{E(\BV)}}   
\providecommand{\gmr}{\ensuremath{g\!-\!r}}%
\providecommand{\umr}{\ensuremath{u\!-\!r}}%
\providecommand{\MH}{\ensuremath{\mbox{[M/H]}}}
\providecommand{\hden}{\ensuremath{n_{\mbox{\scriptsize H}}}}
\providecommand{\loghden}{\ensuremath{\mbox{log}\,n_{\mbox{\scriptsize H}}}}
\providecommand{\Wr}{\ensuremath{\mbox{W}_{r}}}

\shorttitle{}
\shortauthors{Thom~\etal}

\begin{document}

\title{The Gas-Galaxy Connection at $\zabs = 0.35$: \OVI\ and \HI\ Absorption Towards \qso}

\author{C.~Thom\altaffilmark{1,2}, 
J.~K.~Werk\altaffilmark{3}, 
J.~Tumlinson\altaffilmark{2}, 
J.~X.~Prochaska\altaffilmark{3}, 
J.~D.~Meiring\altaffilmark{4},
T.~M.~Tripp\altaffilmark{4},
K.~R.~Sembach\altaffilmark{2}}

\altaffiltext{1}{cthom@stsci.edu}
\altaffiltext{2}{Space Telescope Science Institute, 3700 San Martin Dr, Baltimore MD, 21211, U.S.A}
\altaffiltext{3}{UCO/Lick Observatory, University of California, 1156 High Street, Santa Cruz, CA 95064, USA}
\altaffiltext{4}{Department of Astronomy, University of Massachusetts, 710 North Pleasant Street, Amherst, MA 01003, USA}

\begin{abstract}
  We present observations of \HI\ and \OVI\ absorption systems proximate to a galaxy at $\zgal =
  0.3529$. The absorption was detected serendipitously in {\it Cosmic Origins Spectrograph}
  observations of the low-$z$ QSO \qso ($\zqso = 0.564$). The data show two separate clouds along
  the sightline at an impact parameter of $95\kpc$ from the galaxy. The first is likely
  low-metallicity gas falling onto the galaxy. This assessment is based on the high velocity offset
  of the cloud from the galaxy ($\dv = 365\kms$) and the weak metal line absorption, combined with
  photoionization modeling. The second cloud, with only a modest velocity separation from the galaxy
  ($\dv = 85\kms$), exhibits very strong \OVI\ absorption qualitatively similar to \OVI\ absorption
  seen in the Milky Way halo. Collisional ionization equilibrium models are ruled out by the metal
  line column density ratios. Photoionization modeling implies a length-scale for the \OVI\ cloud of
  $\sim0.1-1.2\Mpc$, which indicates the absorbing gas most likely resides within the local
  filamentary structure. This system emphasizes that kinematic association alone is not sufficient
  to establish a physical connection to galaxies, even at small impact parameters and velocity
  separations.  Observations such as these, connecting galaxies with their gaseous environments, are
  becoming increasingly important for understanding of galaxy evolution and provide constraints for
  cosmological simulations.
\end{abstract}


\section{Introduction}
Two of the most important unanswered questions in galaxy formation concern how galaxies acquire
their gas, and how they recycle it back into their circumgalactic medium (CGM) and the intergalactic
medium (IGM). The starlight of galaxies is relatively easy to observe, and recent surveys such as
the Sloan Digital Sky Survey \citep[SDSS;\ ][]{york-etal-00-SDSS} have yielded uniformly selected,
statistical characterizations of galaxies, such as the luminosity function
\citep{blanton-etal-03-SDSS-LF}, the mass-metallicity relation
\citep{tremonti-etal-04-mass-metallicity} and color-magnitude diagram
\citep{strateva-etal-01-galaxy-CMD,kauffmann-etal-03-SFR}.

By contrast with the stars, the gaseous environments of galaxy halos are poorly understood. \HI\
21\,cm observations are only possible for the closest galaxies, while QSO absorption line studies
have been plagued by small sample sizes, and the ambiguity of associating absorption systems with
the nearby galaxy population. To investigate this crucial aspect of galaxy evolution we have
embarked on a program to empirically characterize gas in the halos of galaxies. Our Cycle 17 Hubble
Space Telescope (HST) large program (PI Tumlinson; PID 11598) will obtain spectra of 39 QSOs, which
were selected to lie behind $L \geq \Ls$ ``target'' galaxies at $z = 0.1 - 0.3$, while our Cycle 18
HST program (PI Tumlinson; PID 12204) will improve coverage of the galaxy luminosity function by
observing 41 sightlines that pass through dwarf galaxy halos. The goal of the survey is to
systematically map the multiphase gaseous structure of galaxy halos on 100\kpc\ scales, with the aim
of unraveling their feedback and accretion processes.

In the Milky Way, gas has been observed in the halo since the early 1960s, both in the 21\,cm
emission of neutral hydrogen \citep{muller-etal-63-HVC-discovery, barnes-etal-01-HIPASS,
  kalberla-etal-05-LAB}, as well as UV and optical absorption lines \citep[e.g.][]{muench-zirin-61,
  vanwoerden-etal-99-nature, tripp-etal-03-ComplexC, sembach-etal-03-OVI-HVC, richter-etal-05-low-density-clumps,
  thom-etal-06-complexWB, thom-etal-08-ComplexC, ben-bekhti-etal-08-CaII,
  collins-etal-09-SiIII-survey}. In order to differentiate halo clouds from disk material, selection
is typically based on the radial velocity of the cloud. Halo clouds with high velocities relative to
the Local Standard of Rest ($\vLSR > 100\kms$) are thus deemed high-velocity clouds
\citep[HVCs;][]{wakker-vanwoerden-97}. Due to this velocity selection, we may be missing some of the
halo clouds when studying the MW as low-velocity halo clouds are hard to distinguish from the disk
\citep{peek-etal-09-low-velocity-HVCs}.

The 21\,cm HVCs trace only the highest column density systems \citep[$\NHI \simgt
10^{18}\cmsq$;][]{putman-etal-03-mshvc, kalberla-etal-05-LAB} and they represent flow of gas
into and out of the disk of the Galaxy. They may provide fresh fuel for star-formation in the
disk, either accreted directly from the IGM, or stripped from dwarf satellite
galaxies. Alternatively, they may be part of outflowing galactic waste, expelled from the disk by
supernova driven winds. Many of these high column density systems lie within $\sim 10-20\kpc$
\citep{vanwoerden-etal-99-nature, thom-etal-06-complexWB, wakker-etal-07-I-distances,
  thom-etal-08-ComplexC,lehner-howk-10-HVC-distances,tripp-song-11-outer-arm}, as determined by
foreground/background stellar probes at known distances. The exception---the Magellanic Stream---is
typically assumed to be at the same distance as the Magellanic Clouds ($\sim 50\kpc$).

Under the current paradigm, the volume between the the cool, dense halo clouds is filled with a
tenuous, low-density, hot halo corona. Its existence was first proposed by \citet{spitzer-56-corona}
as the medium which held the observed cool clouds in pressure equilibrium, and it is a universal
prediction of numerical simulations of \Ls\ galaxies \citep[e.g.\ ][]{white-rees-78-halos,
  keres-etal-09-I-simulation}. For a Milky Way sized halo, accreting gas is expected to be
shock-heated to the virial temperature of $\sim10^6\K$
\citep{birnboim-dekel-03-accretion-shock}. Galactic outflows may also contribute a substantial
amount of hot material \citep{oppenheimer-dave-09-OVI-simulation}.

This hot, low-density gas is difficult to detect directly. Its effect is seen indirectly as a drag
force on clouds moving through it \citep{peek-etal-07-VHVC-distance}, which results in a pronounced
head-tail structure as the denser cool clouds are slowly ablated. This is an effect seen both in
simulations \citep{quilis-moore-01, heitsch-putman-09-simulation} and radio observations
\citep{westmeier-etal-05-effelsberg}. X-ray absorption towards the LMC X-3 X-ray source provides
evidence that the halo gas is hot \citep{wang-etal-05-LMC-X3}. A sample of 25 \OVII\ K$\alpha$
detections also favors an explanation in MW hot halo gas \citep[as opposed to a Local Group
origin;][and references therein]{bregman-lloyd-davies-07-MW-Xray}

Further evidence for this hot halo corona comes from the \OVI\ absorption line in the UV
\citep[][hereafter S03]{sembach-etal-03-OVI-HVC}. \OVI\ traces gas at $\logT \sim 5.5$ in collision
ionization equilibrium \citep{sutherland-dopita-93-gas-cooling}. Using an extensive survey of 102
AGN sightlines passing through the halo, S03 found a covering fraction of $65-85\%$ for column
densities in the range $\logNOVI = 13 - 14.5$. Some of the Milky Way \OVI\ is associated with the
\HI\ HVCs and likely arises in interface layers between cool and hot gas.  Some detections, however,
were not correlated with detectable \HI\ absorption or emission, and favor an extended ($R \simgt
70\kpc$), hot ($\mbox{T} > 10^{6}\K$) halo. This conclusion was based on the column density ratios
of \OVI\ to other highly ionized species (such as \NV\ and \CIV), as well as a strong correlation
between \NOVI\ and the line widths (S03).

In QSO sightlines probing the IGM and low-$z$ galaxy halos, the physical conditions of \OVI\
absorbers are harder to determine, as it is harder to distinguish between collisional ionization in
hot gas, and cool gas photoionized by the meta-galactic UV background field. \OVI\ can be found in a
diverse range of environments, from galaxies to groups, filaments and the IGM
\citep{savage-etal-98-H1821+643, tripp-savage-00-PG0953+415_z0.1423,
  tripp-etal-01-H1821+643_z0.1212, prochaska-etal-06-PKS0405-12_galaxies,
  stocke-etal-06-OVI-galaxies, lehner-etal-09-PKS0312-z0.2028, prochaska-etal-11-WFCCD-OVI}. Several
studies have shown a correlation with galaxies on large spatial scales
\citep[e.g.][]{stocke-etal-06-OVI-galaxies, prochaska-etal-11-WFCCD-OVI}, and there appears to be a
correlation with emission-line dominated galaxies \citep{chen-mulchaey-09-I-survey}. Several recent
statistical surveys have studied \OVI\ in the low-$z$ IGM with the Space Telescope Imaging
Spectrograph (STIS) \citep{tripp-etal-08-OVI, danforth-shull-08-OVI, thom-chen-08-I-OVI-statistics},
but while some of the gas can be shown to be cool \citep{tripp-etal-06-PG0953+593-galaxies,
  thom-chen-08-II-OVI-absorbers}, much of it is multi-phase gas or its ionization state cannot be
determined \citep[e.g.][]{tripp-etal-08-OVI, thom-chen-08-II-OVI-absorbers}. \OVI\ is thus a useful
gas diagnostic, but is sometimes ambiguous, and does not correspond to a unique environment. The
installation of the Cosmic Origins Spectrograph \citep[COS;][]{green-etal-03-COS} on-board HST has
dramatically opened up the number of QSOs available for detailed studies of \OVI\ in the low-$z$ IGM
and galaxy halos, and several recent results have reported analyses of individual low-$z$ \OVI\
systems \citep{savage-etal-10-PKS0405-z0.167, danforth-etal-10-1ES1553, savage-etal-11-HE0153,
  yao-etal-11-model}.

Here we report the serendipitous detection of halo \OVI\ absorption and metal-poor gas in or near a
galaxy halo at $z = 0.35$. This galaxy was not the targeted galaxy, and thus the detections are
truly serendipitous; initial statistical analysis of all the survey targets will be presented in a
forthcoming series of publications (Tumlinson \etal, in prep). The absorbers are detected at an
impact parameter of $\rho = 95\kpc$. The \OVI\ cloud may be tracing a photoionized filamentary
structure near the galaxy, while the metal-poor cloud is more likely to be tracing gas infall into
the galaxy halo. Companion papers by \citet{tumlinson-etal-11-J1009-LLS} and
\citet{meiring-etal-11-J1009-DLA} present other first results from our Cycle 17 halos survey,
including the observation of the strongest \OVI\ absorber seen in the low-$z$ IGM, and initial
results on the first sample of damped \Lya\ absorbers detected with COS (respectively).  In
Section~\ref{sec: data} we describe the observations and data reduction. Section~\ref{sec: results}
gives the main results, and Section~\ref{sec: discussion} discusses the implications for the Milky
Way and galaxies in general.  When necessary, we adopt a flat cosmology with $\Omegam = 0.27$,
$\Omegal = 0.73$ and $\Hn = 71\kmsMpc$.

\section{Data}
\label{sec: data}
\subsection{COS Spectroscopy}
\label{subsec: COS}

\begin{deluxetable}{llclr}
\tabletypesize{\scriptsize}
\tablecaption{\label{tab: journal_observations}Summary of the COS Observations}
\tablewidth{0pt}
\tablehead{
  \colhead{Dataset} &
  \colhead{Grating} & 
  \colhead{Cenwave}&
  \colhead{Coverage} &
  \colhead{$t_{\rm exp}$}\\
  \colhead{} &
  \colhead{} &
  \colhead{(\A)} &
  \colhead{(\A)} &
  \colhead{(\mbox{s})} 
}
\startdata
lb5n42qgq  &  G130M &  1291 & $1132 - 1433$  &  2101 \\
lb5n42qlq  &  G130M &  1309 & $1153 - 1440$  &  1561 \\
lb5n42qqq  &  G160M &  1600 & $1410 - 1772$  &  1177 \\
lb5n42qsq  &  G160M &  1600 & $1410 - 1772$  &   804 \\
lb5n42qvq  &  G160M &  1623 & $1433 - 1796$  &  1964
\enddata
\tablecomments{Journal of COS Observations of \qsofull\ at $\zqso = 0.564$.}
\end{deluxetable}

The UV spectrum of the quasar SDSS J094331.61+053131.4 (hereafter \qso) was obtained with COS on 06
March 2010. The raw data were processed through the standard COS reduction and calibration pipeline,
{\it calcos} (v2.12). More detailed information can be found in the COS instrument
handbook\footnote{http://www.stsci.edu/hst/cos/documents/handbooks/}. We obtained spectra with both
medium resolution FUV gratings, G130M and G160M. Table~\ref{tab: journal_observations} shows a
journal of the observations, grating settings, and exposure times.

We developed software to align and co-add the individual, calibrated x1d spectra output by {\it
  calcos}. Since much of our data is in the low count regime ($N<30$), we operated directly on
photon counts, rather than flux calibrated data. This allowed us to correctly estimate the Poisson
errors associated with the number of counts in a given pixel
\citep{gehrels-etal-86-small-number-statistics}. Each exposure with a given grating and central
wavelength setting was co-aligned using common Galactic absorption lines to determine the relative
shifts between the individual exposures. The exposures were then shifted to a common reference and
co-added. This was process was performed individually for each detector segment. The two detector
segments for each grating were then aligned with each other using absorption lines of similar
strengths (e.g.\ \SiIII\,1206 vs \CII\,1334). Finally, the G130M and G160M spectra were co-aligned
and co-added using lines of the same species (e.g.\ \SiII\,1260 vs \SiII\,1526).

Co-addition was performed by a simple sum of counts, after alignment. For each pixel, we tracked the
wavelength, number of counts, upper and lower error estimates due to Poisson statistics, and the
effective exposure time. We applied a flat-field correction by taking the STScI COS team 1d flats
(D. Massa, priv. comm.), processed with a low-pass filter, to remove the high-frequency noise. These
flats account for the grid-wire pattern in the COS FUV detectors, and are applied to the effective
exposure time for each pixel (note that they are not applied to the counts directly, since we use
the counts to calculated the Poisson errors). Since we accumulate counts, the spectra contain sharp
discontinuities in regions where the wavelength dithering resulted in a larger effective exposure
time, or in regions affected by the grid-wire pattern. Thus our counts spectra do not have a smooth
continuum but the count {\it rate} spectrum does.

Finally, we normalized the count-rate spectrum by piecewise fitting of low-order Chebyshev
polynomials to chunks of spectrum $\sim 50-100\A$ long. Our procedure yields a normalized spectrum
with both upper and lower error estimates. In regions of high counts ($\simgt 30$), these estimates
converge to the usual gaussian approximation of $\sqrt{N}$, with the upper and lower estimates the
same (i.e. converging to the usual 1$\sigma$ errors). To be conservative, when quoting errors on
measured quantities, we quote the larger of these two error estimates.

The S/N of the resulting spectra varies from $\sim 4$ per resolution element\footnote{The COS
  detector and calcos oversamples the resolution element with 6 detector ``pixels'', which are
  defined by the clock resolution and electronics of the detector.  We bin the resulting spectra by
  3 pixels. This retains Nyquist sampling of the resolution element, but improves the S/N.} at the
shortest wavelengths ($\sim 1135\A$), peaking at $\sim 8$ per resel at $\sim 1400\A$, and declining
to $\sim 2-3$ per resel at the red end of the spectrum ($\sim 1800\A$). By choosing multiple
appropriate central wavelength settings for each grating, we ensured continuous wavelength coverage
(i.e. no gaps), at the expense of lower S/N in the wavelength regions with only one grating setting.

\subsection{Keck HIRES Spectroscopy}
We obtained a high resolution optical spectrum of \qso\ on 26 March 2010 using the High Resolution
Echelle Spectrometer \citep[HIRES;][]{vogt-etal-94-HIRES} on Keck I. We employed the UV
cross-disperser and the C1 decker (slit width 0.86\arcsec), resulting in a resolution of $\sim
48000$ or $\sim 6\kms$. The grating angles were set to give coverage down to $3050\A$, in order to
detect low-$z$ \MgII\ absorption. The data were reduced with the HIRedux pipeline included with the
XIDL package\footnote{http://www.ucolick.org/$\sim$xavier/HIRedux/index.html}. Individual orders
were normalized after extraction with a series of Chebyshev polynomials to remove the echelle blaze
function, and then combined into a final 1D spectrum. 

\subsection{Galaxy Spectroscopy}
After the COS observations were conducted we obtained spectra of several galaxies in the field close
to the QSO with the Low Resolution Imaging Spectrometer \citep[LRIS;][]{oke-etal-95-LRIS} on the
Keck I telescope. We obtained long-slit spectra of 4 galaxies on 25 March 2010. We employed the
5600\A\ dichroic, with the 600/7500 grating on the red side, and the 600/4000 grism on the blue
side. The slit size was 1\arcsec, yielding a FWHM resolution of $4.7\A$ ($\sim 200\kms$) over a
wavelength range of $5600 - 8200\A$ on the red side; the blue side covered the range $3000 - 5500\A$
at a resolution of $3.9-4.1\A$ FWHM ($\sim 300\kms$).  The data were reduced using the LRIS pipeline
in the XIDL package.

\begin{figure}
  \epsscale{1.2}
  \plotone{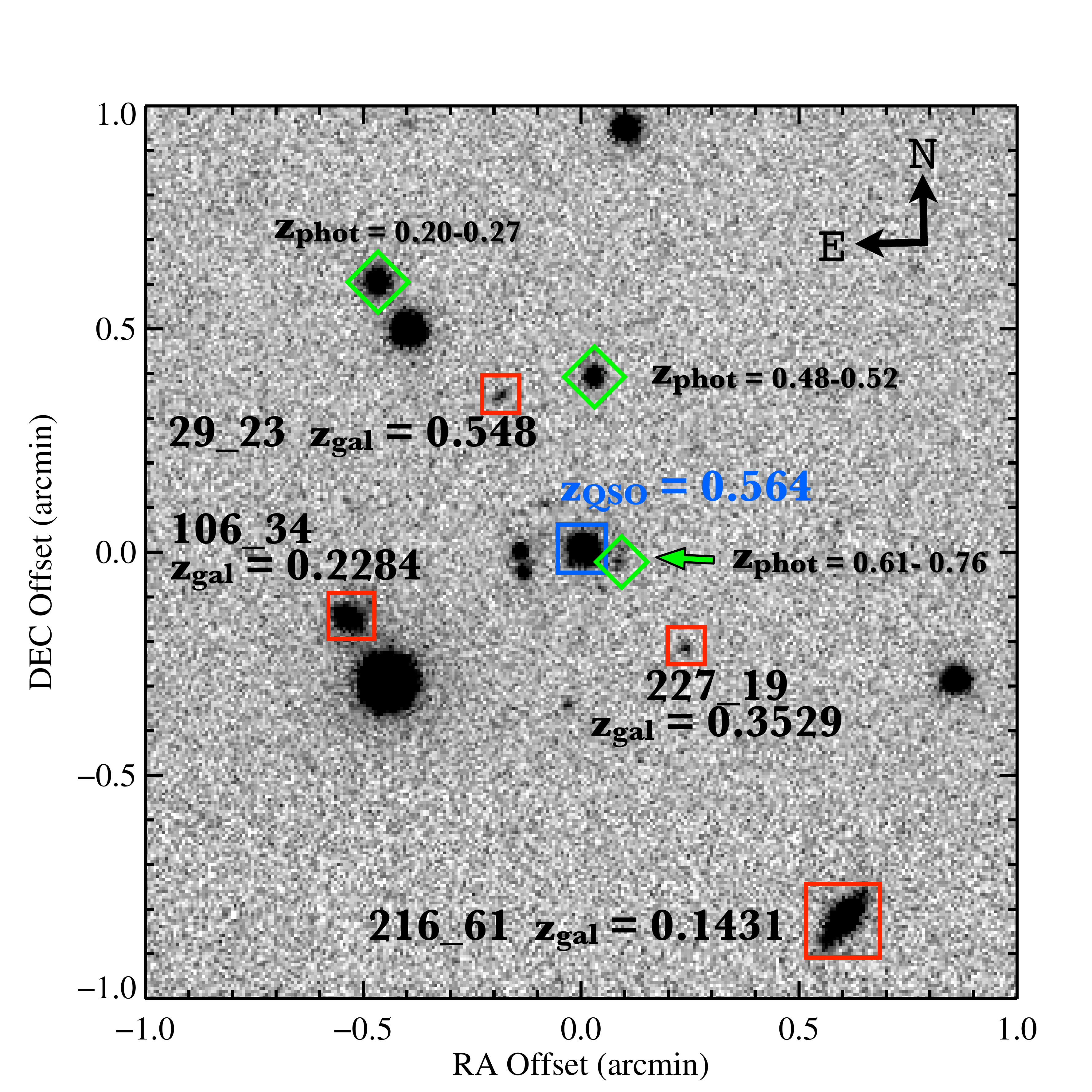}
  \caption{SDSS {\it i}-band image of the field around \qso. The QSO is centered and surrounded with
    a blue box. Galaxies are labeled with a polar co-ordinate system based on the position angle
    (degrees east of north) and the angular separation from the QSO (arcsec). Galaxies with
    spectroscopic redshifts are enclosed in red boxes, while galaxies with only photometric
    estimates are enclosed in green diamonds. Objects not labeled are stars.}
  \label{fig: keck_field}
\end{figure}

Fig~\ref{fig: keck_field} shows an image of the field taken from the SDSS, centered on the QSO
position. Objects classified as galaxies in SDSS are labeled in a polar co-ordinate system, with a
position angle (degrees east of north) and an angular distance from the QSO (in arcsec). Galaxies
for which we obtained spectroscopic redshifts are enclosed in red boxes, and the measured redshifts
included in the label. Objects classified as galaxies in SDSS but which lack spectroscopic redshifts
are enclosed in green diamonds. The maximum range of the SDSS photometric redshift estimates are
listed. We have also examined the photometric redshift probability distributions for these objects
produced by \citet{cunha-etal-09-SDSS-photoz}. None of these objects are consistent with being at
the redshift of the absorbers, although we note that contamination of objects very close to the QSO
by the QSO light may be an issue, and it would be ideal to obtain further spectroscopy follow-up for
these galaxies. Unlabeled objects in Figure~\ref{fig: keck_field} are classified as stars. Higher
resolution imaging with e.g. HST, is an ongoing component of our multi-instrument project, and would
be helpful in the future for quantitative measures of galaxy morphology.

\section{Results}
\label{sec: results}
\subsection{Absorption Lines}
\begin{figure}
  \epsscale{1.2}
  \plotone{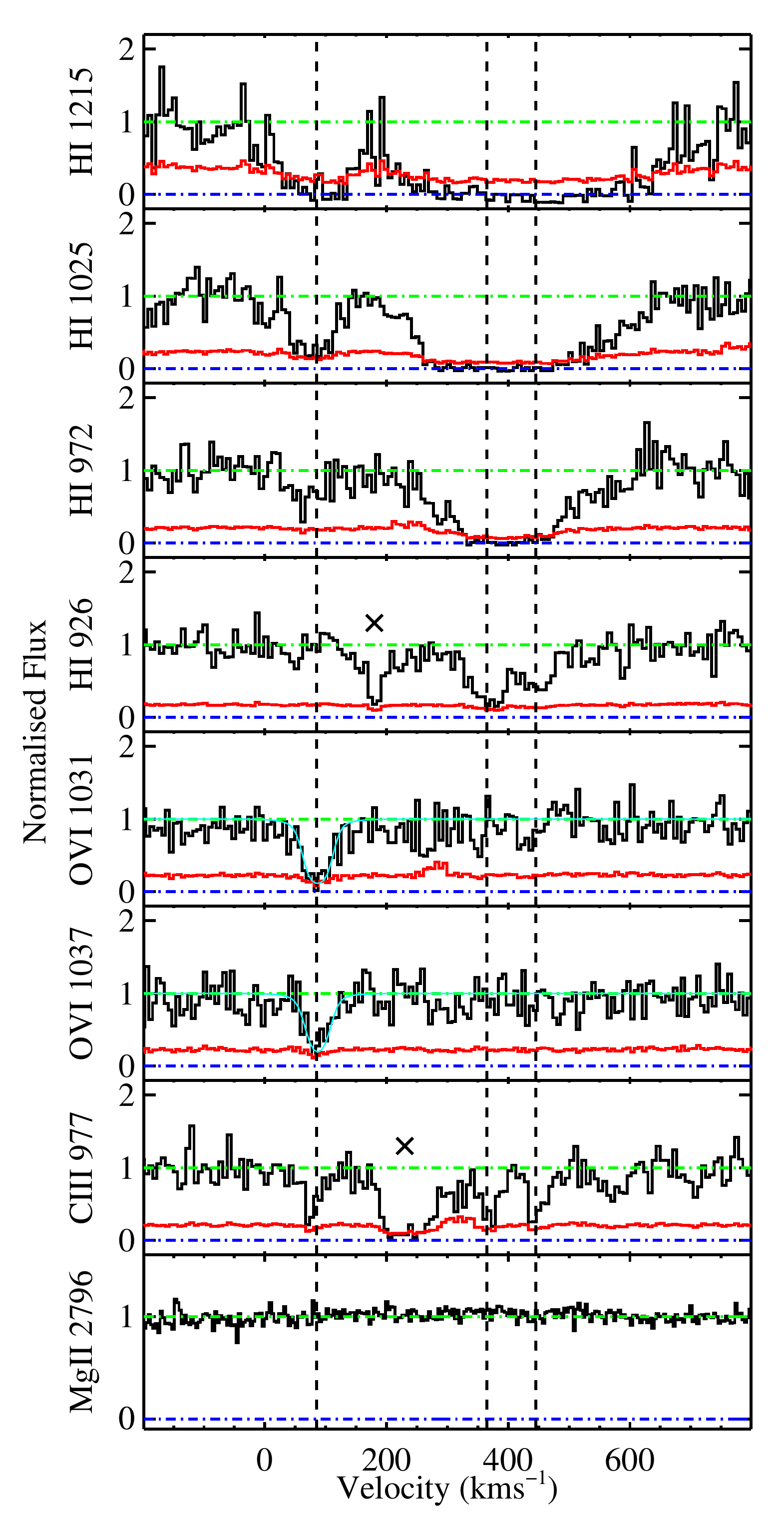}
  \caption{Velocity stack of detected lines in the \OVI\ and metal-poor clouds.  Strong absorption
    \HI, \OVI, and \CIII\ absorption can be seen at $v = 85\kms$, while the much strong \HI\
    absorption and weak \CIII\ absorption at $v = +365, 445\kms$ is due to the metal-poor cloud.
    \MgII\ is not detected in either the \OVI\ or metal-poor cloud. Velocities are with respect to
    the systemic redshift of \galid, $\zgal = 0.3529$. Small crosses mark absorption at other
    redshifts. Profile fits to the \OVI\ absorption are shown overplotted in thin cyan lines. See
    below for details.}
  \label{fig: stack-plot}
\end{figure}

In Figure~\ref{fig: stack-plot} we show the absorption lines affiliated with the galaxy
\galid. We plot the continuum normalized spectra in a velocity space, with respect to the
spectroscopic redshift of the galaxy, $\zgal = 0.3529$ (all velocities in this paper are with 
respect to this zero-point). 

At $v = +365\kms$ we detect strong saturated \Lya\ absorption. The first 5 \HI\ lines are saturated,
and the system breaks into two main components in the higher order Lyman lines, at velocities of $v
= +365, +445\kms$. Both components have associated weak \CIII\ absorption. The spectrum also shows a
hint of possible \OVI\ absorption, but these features are only significant at the $2\sigma$ level,
and we consider them non-detections. Higher quality data would be required to confirm these
features. Significantly, we detect no \MgII\ absorption in the HIRES data to very sensitive upper
limits. Due to its lack of strong metal line absorption, we describe this cloud as ``metal-poor'', a
name we will justify in Section~\ref{subsec: metal-poor}. This system also contains a much weaker
\HI\ component at $+285\kms$ with no associated metal absorption. Figure~\ref{fig: stack-plot} shows
the first 3 Lyman series lines, as well as the relatively weak \HI\,926 line, which shows the
component structure.

At a lower velocity to the metal-poor cloud, we clearly detect \HI, \OVI\ and \CIII\ absorption at
$v = +85$. Due to the obvious detection of \OVI\ absorption, we label this absorber the ``\OVI\
cloud''. It is immediately apparent that the \NHI\ is significantly lower in the \OVI\ cloud than
the metal-poor cloud, while the \CIII\ absorption strength is the same in both clouds. While \OVI\
is lacking in the metal-poor cloud, it is clearly detected here. The HIRES spectra show no detection
of \MgII\ in either cloud to sensitive limits (see Table~\ref{tab: column_densities}).

We measured rest-frame equivalent widths of the absorption lines ($W_r$) by direct integration of
the data, and the results are given in Table~\ref{tab: column_densities}. To be conservative, we
adopted the larger of the two Poisson error estimates in calculating equivalent width errors (see
Section~\ref{sec: data}). We converted equivalent widths to column densities assuming the gas is
optically thin (i.e. linear curve of growth) except where noted in the following section; saturated
lines are given as lower limits.

\begin{deluxetable}{lrrll}
\tabletypesize{\scriptsize}
\tablecaption{\label{tab: column_densities}Measured column densities}
\tablewidth{0pt}
\tablehead{
  \colhead{Line} & 
  \colhead{Velocity} & 
  \colhead{W$_r$} & 
  \colhead{$N$} & 
  \colhead{Notes} \\
  \colhead{} &
  \colhead{[\kms]}&
  \colhead{[\mA]}&
  \colhead{[log \cmsq]}&
  \colhead{}\\
}
\startdata
\sidehead{Metal-poor cloud}
\HI\,1215        &  $285$  &  \nodata        &  \nodata          & Inseparable   \\
\HI\,1025        &  $285$  &  \nodata        &  \nodata          & Inseparable   \\
\HI\,972         &  $285$  &  $  97 \pm 14$  &  $14.6 \pm  0.1$  & \\
\HI\,949         &  $285$  &  $  81 \pm 23$  &  $14.9 \pm  0.1$  & \\
\HI\,937         &  $285$  &  $  72 \pm 12$  &  $15.1 \pm  0.1$  & \\
\HI\,1215        &  $365$  &  $1765 \pm 74$  &  $>14.5 \pm  0.1$  & Saturated \\
\HI\,1025        &  $365$  &  $1158 \pm 35$  &  $>15.2 \pm  0.1$  & Saturated \\
\HI\,972         &  $365$  &  $ 586 \pm 17$  &  $>15.4 \pm  0.1$  & Saturated \\
\HI\,949         &  $365$  &  $ 495 \pm 28$  &  $>15.6 \pm  0.1$  & Saturated \\
\HI\,937         &  $365$  &  $ 214 \pm 10$  &  $>15.5 \pm  0.1$  & Saturated \\
\HI\,930         &  $365$  &  \nodata        &  \nodata           & Blended \\
\HI\,926         &  $365$  &  $ 169 \pm 11$  &  $15.8 \pm  0.1$   & \\
\HI\,923         &  $365$  &  \nodata        &  \nodata           & Blended \\
\HI\,920         &  $365$  &  $  95 \pm 13$  &  $15.9 \pm  0.1$   & \\
\HI\,919         &  $365$  &  $  77 \pm 12$  &  $15.9 \pm  0.1$   & \\
\CIII\,977       &  $365$  &  $  72 \pm 10$  &  $13.0 \pm  0.1$   & \\
\CII\,1036       &  $365$  &  $     \pm 12$  &  $<13.0        $   & Non-detection\\
\OVI\,1031       &  $365$  &  $  34 \pm 16$  &  $<13.2$           & Non-detection\\
\SiIII\,1206     &  $365$  &  $     \pm 39$  &  $<12.3$           & Non-detection\\
\MgII\,2796      &  $365$  &  $     \pm  5$  &  $<11.1$           & Non-detection\\
\HI\,1215        &  $445$  &  \nodata        &  \nodata           & Inseparable   \\
\HI\,1025        &  $445$  &  \nodata        &  \nodata           & Inseparable   \\
\HI\,972         &  $445$  &  \nodata        &  \nodata           & Inseparable   \\
\HI\,949         &  $445$  &  \nodata        &  \nodata           & Inseparable   \\
\HI\,937         &  $445$  &  $ 187 \pm 11$  &  $>15.5$           & Saturated \\
\HI\,930         &  $445$  &  $ 140 \pm 11$  &  $15.6 \pm  0.1$   & Blended? \\
\HI\,926         &  $445$  &  $ 102 \pm 12$  &  $15.6 \pm  0.1$   & \\
\HI\,923         &  $445$  &  \nodata        &  \nodata           & Blended  \\
\HI\,920         &  $445$  &  $  87 \pm 14$  &  $15.9 \pm  0.1$   & \\
\HI\,919         &  $445$  &  $  39 \pm 11$  &  $15.6 \pm  0.1$   & \\
\CIII\,977       &  $445$  &  $  80 \pm 13$  &  $13.1 \pm  0.1$   & \\
\OVI\,1031       &  $445$  &  $  36 \pm 17$  &  $<13.1$           & Non-detection \\
\SiIII\,1206     &  $445$  &  $     \pm 33$  &  $>12.2$           & Non-detection \\
\hline
\sidehead{\OVI\ Cloud} 
\HI\,1215        &  $ 85$  &  $ 530 \pm 47$  &  $>14.0$         & Saturated \\
\HI\,1025        &  $ 85$  &  $ 206 \pm 18$  &  $14.4 \pm  0.1$ & Saturated \\
\HI\,972         &  $ 85$  &  $  80 \pm 16$  &  $14.5 \pm  0.1$ & \\
\HI\,949         &  $ 85$  &  $  41 \pm 28$  &  $14.6 \pm  0.4$ & Non-detection\\
\OVI\,1031       &  $ 85$  &  $ 150 \pm 14$  &  $14.1 \pm  0.1$ & \\
\OVI\,1037       &  $ 85$  &  $ 107 \pm 15$  &  $14.2 \pm  0.1$ &\\
\CIII\,977       &  $ 85$  &  $  71 \pm 14$  &  $13.0 \pm  0.1$ &\\
\CII\,1036       &  $ 85$  &  $     \pm 26$  &  $<13.4$         & Non-detection\\
\SiIII\,1206     &  $ 85$  &  $     \pm 42$  &  $<12.4$         & Non-detection\\
\MgII\,2796      &  $ 85$  &  $     \pm  4$  &  $<11.0$         & Non-detection\\
\enddata
\tablecomments{Measured column densities for the metal-poor halo cloud at $v=+365\kms$, and the
  \OVI\ cloud at $v = 85\kms$. Rest-frame equivalent widths are measured directly. Column densities
  are calculated assuming a linear curve-of-growth; lower limits are indicated for saturated lines,
  and $1\sigma$ upper limits for lines not detected. \Wr\ is given as $\pm 1\,\sigma$ errors for
  non-detections. Cases of obvious blending with galactic absorption or other intervening systems
  are not reported.}
\end{deluxetable}

\begin{figure}
  \epsscale{2.4}
  \plottwo{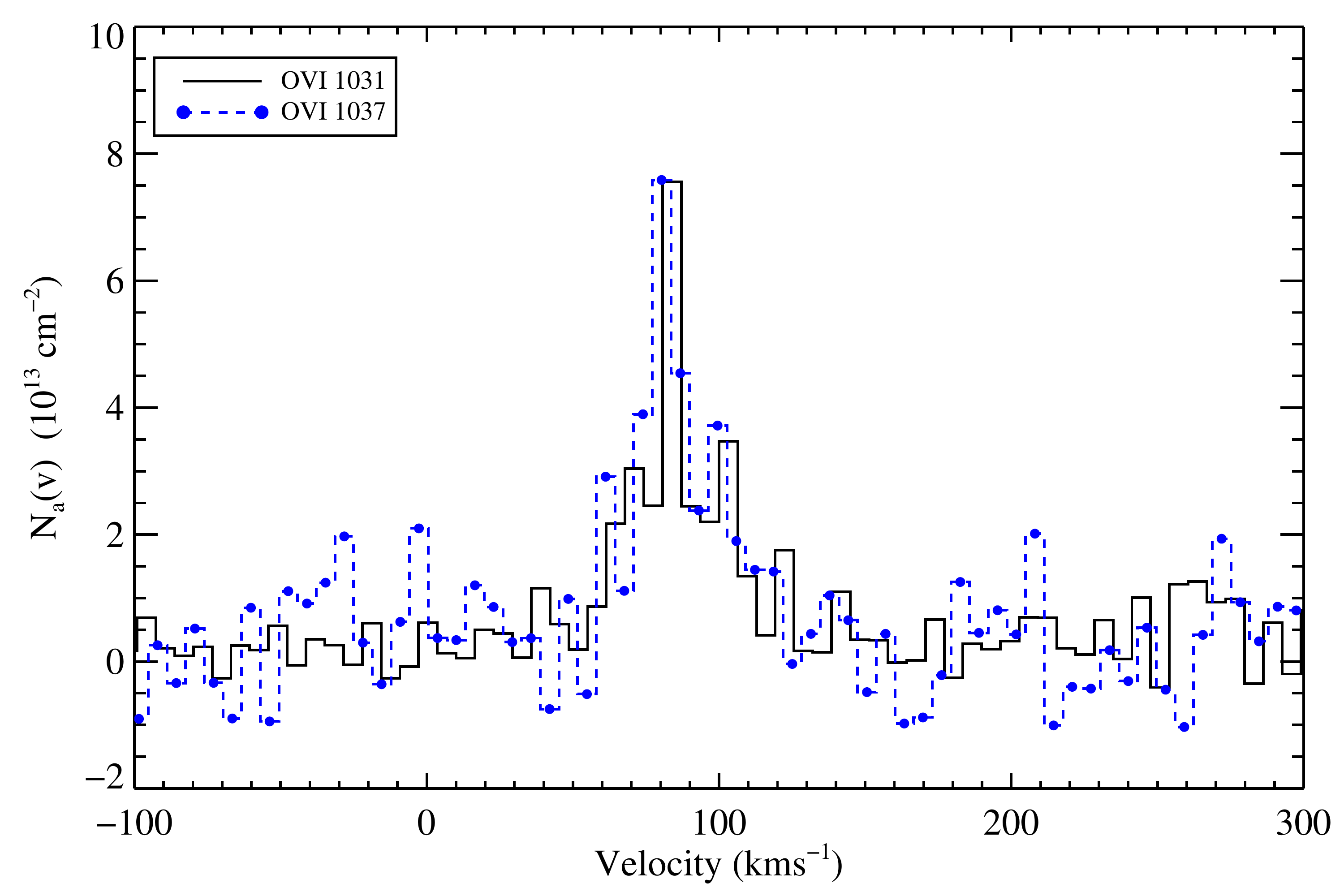}{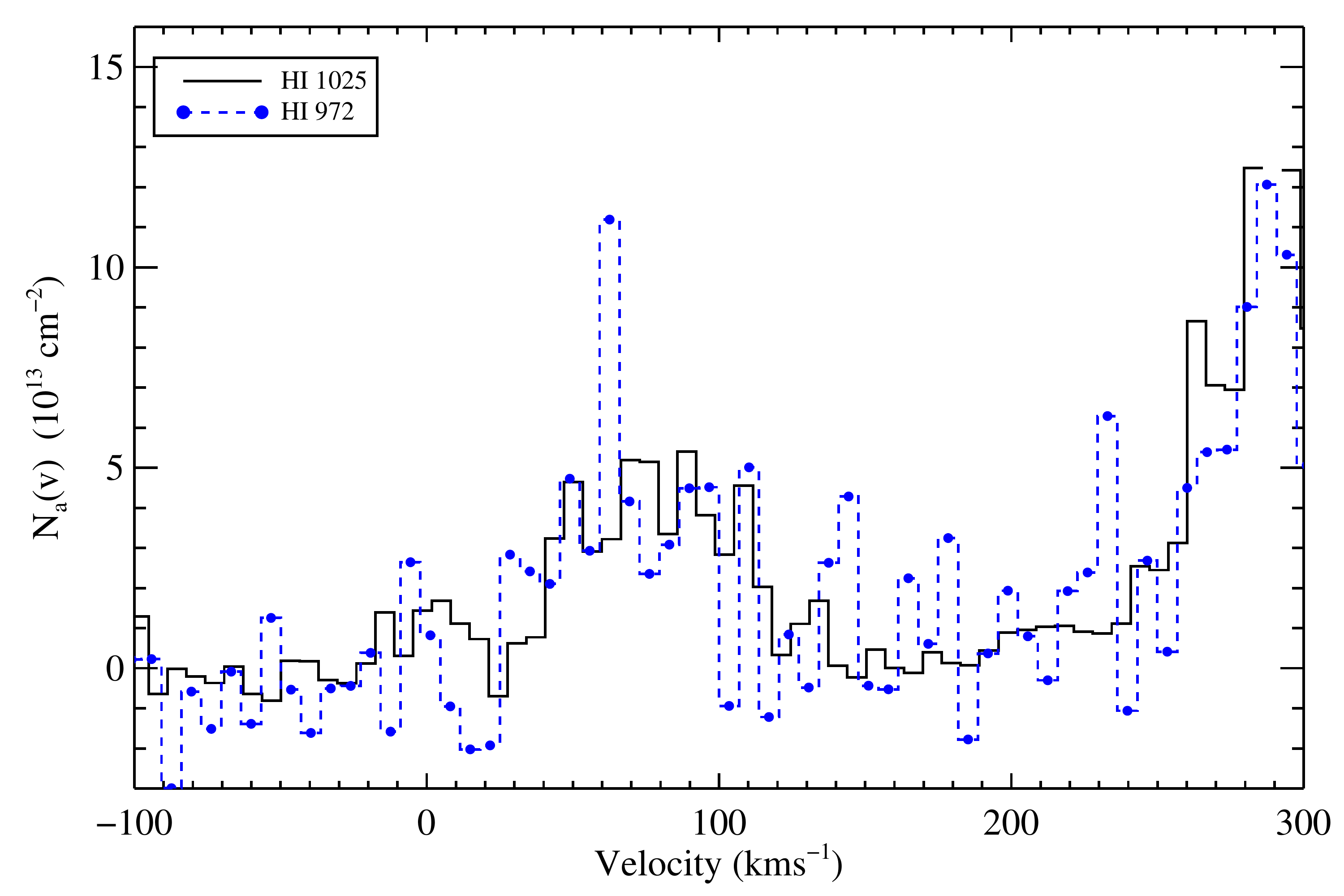}
  \caption{Comparison of apparent column density profiles. The top panel shows both members of the
    \OVI\ doublet; the bottom panel shows the \Lyb\ and \Lyc\ \HI\ lines. Both sets of profiles are
    well matched. The \OVI\,1037 line shows an anomalous noise spike, which is apparent in the top
    blue (dashed) curve. In the bottom panel, the edge of the metal-poor cloud can be seen at $\sim 250\kms$.}
  \label{fig: OVI-AOD}
\end{figure}

For the \OVI\ and unsaturated \HI\ lines we generated apparent column density profiles
\citep{savage-sembach-91-AOD}. The apparent column density is defined as the observed normalized
flux decrement in a given bin converted to a column density using the relation $N_a(v) =
(3.767\times10^{14}\,\tau_a(v))/(f\,\lambda)$ where $\tau_a(v) = ln[I_c/I]$ is the apparent optical
depth, $f$ is the oscillator strength of the transition, and $\lambda$ is measured in $\A$. Note
that the continuum intensity $I_c$ is unity for a continuum-normalized spectrum. The \OVI\ and \HI\
apparent column density profiles are shown in Fig~\ref{fig: OVI-AOD}.

\subsection{Voigt Profile Fits}
Since the metal lines come close to zero flux in several cases, we also performed Voigt profile
fitting to check for unresolved saturation. Fitting was performed on the \OVI\ and \CIII\ lines in
both clouds, and results are shown in Table~\ref{tab: profile_fitting}. 
\newline

{\bf Metal-poor Cloud}\\
We fit the \CIII\ line in both strong components of the metal-poor cloud. The profile fits yield a
higher column density ($\sim 0.4\dex$) than is obtained in the optically thin case, but with
correspondingly larger errors ($\sim 0.3\dex$), such that the two measurements are in statistical
agreement.  
\newline

{\bf \OVI\ Cloud}\\
The \OVI\ apparent column density curves of the \OVI\ cloud shown in Figure~\ref{fig: OVI-AOD} show
some evidence of asymmetry in the \OVI\ profile, indicating that a two-component fit may be
justified for the \OVI. We performed fits using both a single and a double-component model, using
two independant fitting codes, and taking into account the tabulated COS line-spread functions, but
the S/N was not sufficient to distinguish between the two cases. This is reflected in the large
parameter errors for the two component fit. The total column density is the same in with 1 or 2
components, and is $\sim 0.4\dex$ higher than was obtained above by assuming the gas is optically
thin. It is therefore likely that we are seeing a mild level of unresolved saturation. Since the
profile fitting is more robust to saturation in the line core, it is a more reliable measure of the
column density in the system. Moreover, the profile fits do not formally reach zero flux (see
Figure~\ref{fig: stack-plot}), nor do the data show strong evidence of saturation, which leads us to
believe that the unresolved saturation is mild, and we can confidently proceed by adopting the
fitting results as our final \OVI\ column densities.

The \CIII\ column density in the \OVI\ cloud was also $\sim 0.4\dex$ higher than predicted by the
optically thin case, but the fitting errors were very large ($0.6-0.7\dex$). Thus the \CIII\ column
density in the \OVI\ cloud is poorly constrained by the profile fitting, because the line is at best
marginally resolved. This is shown clearly by the low doppler $b$-value, and correspondingly large
errors in both $N$ and $b$. This uncertainty in the \CIII\ column density is the dominant source of
uncertainty in our analysis of the \OVI\ cloud below. We also note that there is an 8\kms offset in
the centroids of the \OVI\ and \CIII\ lines in the \OVI\ cloud as a result of errors in the COS
wavelength calibration.

\begin{deluxetable}{llll}
\tabletypesize{\scriptsize}
\tablecaption{\label{tab: profile_fitting}\OVI\ Voigt profile fitting results}
\tablewidth{0pt}
\tablehead{
  \colhead{Line} & 
  \colhead{$v$} &
  \colhead{$N$} & 
  \colhead{$b$} \\ 
  \colhead{} &
  \colhead{[\kms]} &
  \colhead{[\cmsq]}&
  \colhead{[\kms]}\\
}
\startdata
\sidehead{Metal Poor Cloud}
\CIII    & $369 \pm 3$  & $13.5 \pm 0.2$   &  $13 \pm 7$      \\
\CIII    & $447 \pm 3$  & $13.5 \pm 0.3$   &  $12 \pm 7$      \\
\hline
\sidehead{\OVI\ Cloud---Single Component}
\OVI      & $87 \pm  2$ &  $14.6 \pm 0.1$  &  $17 \pm 3$    \\
\CIII     & $76 \pm  3$  & $13.5 \pm 0.7$  &  $ 9 \pm 8$    \\
\sidehead{\OVI\ Cloud---Two Components}
\OVI      & $84 \pm  4$  & $14.5 \pm 0.2$  &  $14 \pm 8 $   \\
\nodata   & $92 \pm 12$  & $14.2 \pm 0.3$  &  $43 \pm 23$   \\

\enddata
\tablecomments{Results of Voigt profile fits to metal lines.} 

\end{deluxetable}

\subsection{Galaxy Star-Formation Rate and Metallicity}
\begin{figure}
  \epsscale{1.2}
  \plotone{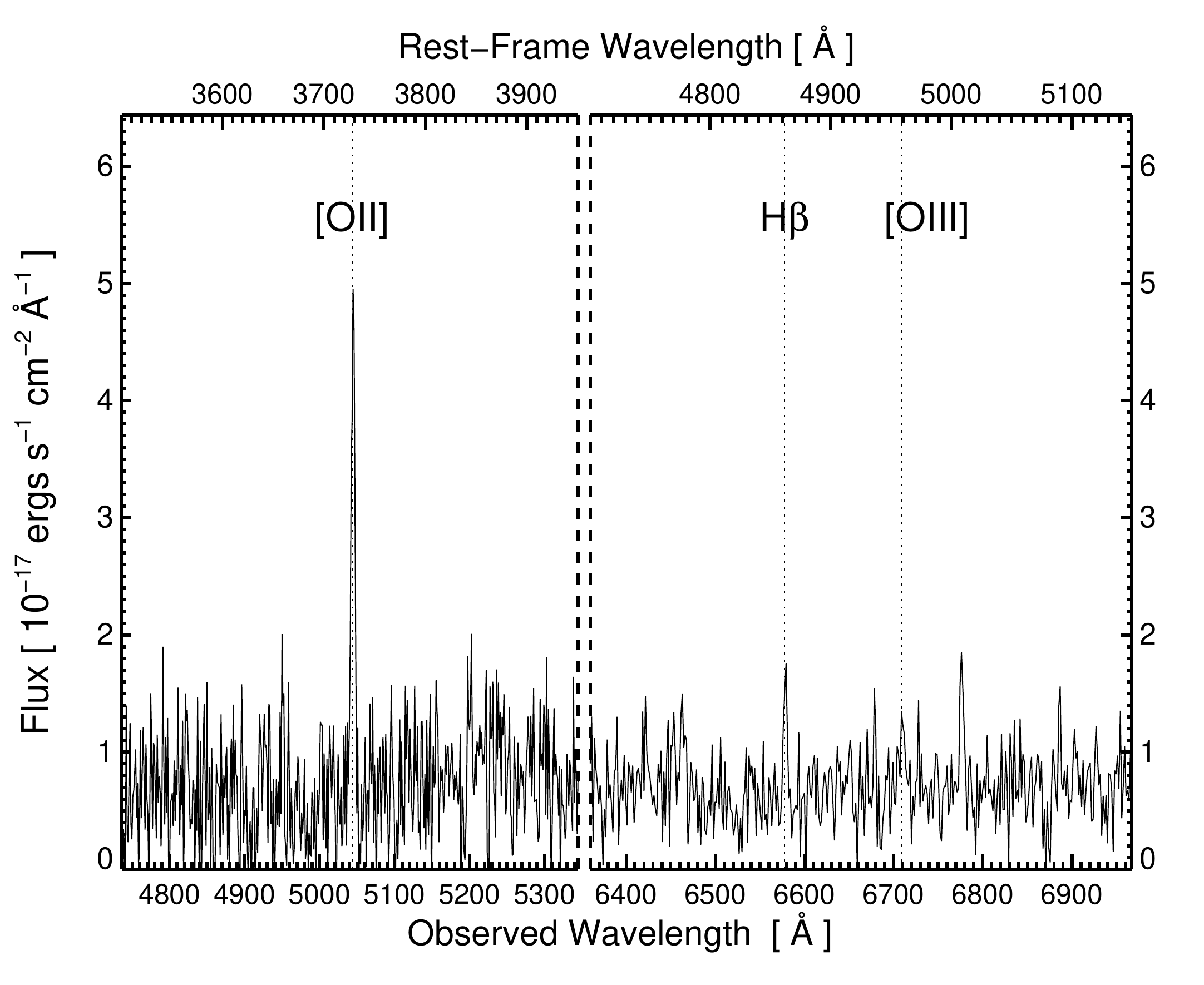}
  \caption{Keck/LRIS spectrum of galaxy \galid. Weak emission lines of [\OII]\,3727, \Hb\ and
    [\OIII]\,5007 can be seen, placing the galaxy at a redshift of $\zgal = 0.3529$.}
  \label{fig: gal_spec}
\end{figure}

We determined the galaxy redshifts by fitting template spectra to the LRIS spectra. This template
fitting is a modified version of the SDSS {\it zfind} code, which operates by fitting a linear
combination of SDSS QSO eigenspectra to the galaxy spectrum \citep[see][for
details]{werk-etal-11-galaxies}.  Galaxy \galid, at $\zgal = 0.3529 \pm 0.0001$, is the only galaxy
within 1\arcmin\ which is spectroscopically confirmed to be at the same redshift as the
absorbers. In all cases, we use the redshift of the galaxy to set the systemic redshift, and measure
velocity offsets from this redshift. The formal fitting error on the redshift is much less than the
quoted redshift uncertainty, but systematic effects (e.g. in the wavelength calibration due to
spectrograph flexure) limit the accuracy to $\sim 25\kms$, which we adopt as our redshift
uncertainty. We recall, for comparison, that the COS resolution is $15-20\kms$. The galaxy spectrum
shows a few, very weak emission lines and the S/N is not sufficient to detect any
absorption. Despite the relative weakness of the emission lines, the redshift determination is
confirmed by the detection of [\OII], [\OIII] and \Hb. The spectrum of \galid\ is shown in
Figure~\ref{fig: gal_spec}.

Since any star-formation rate (SFR) and metallicity indicators we derive depend sensitively on the
measured line strength, care is needed when flux-calibrating the galaxy spectra. We attempt to
ensure consistent flux calibration of both blue and red sides of the LRIS spectra. The seeing during
the LRIS observations was not exceptional, so light losses from a 1\arcsec\ slit were corrected by
comparing our spectra with SDSS photometry. We accomplish this absolute flux correction by
convolving the LRIS spectra with SDSS $ugriz$ filter response curves \citep[see][for a description
of the IDL code ``spec2mag'']{da-silva-etal-11}, and comparing the observed spectral apparent
magnitudes with the SDSS catalog apparent magnitudes \citep[see][for
details]{werk-etal-11-galaxies}. This process yields an independent multiplicative correction for
both blue and red spectra. We derive flux correction factors of 2.2 for the blue side (0.88\,mag in
{\it g}-band) and 2.0 for the red side (0.75\,mag in {\it r}-band). Note that this corresponds to
slit losses of some $50-55\%$, and emphasizes the importance and difficulty of absolute flux
calibration.

We took de-reddened magnitudes and colors for \galid\ from SDSS, and computed the distance modulus
and absolute magnitude with the {\it kcorrect} software \citep{blanton-roweis-07-kcorrect} using the
measured redshift, all of which are listed in Table~\ref{tab: galaxy_properties}. For comparison, we
used the SDSS luminosity function parameters from \citet{blanton-etal-03-SDSS-LF}. These parameters
are strictly appropriate only for comparisons at $z = 0.1$ so for the calculation of the luminosity
in Table~\ref{tab: galaxy_properties} we correct to $z = 0.1$; in all other cases we correct to $z =
0$.  We also used the \Hb\ emission line to derive a star-formation rate (SFR). \Ha\ is redshifted
out of the LRIS wavelength range, but \Hb\ is detected. We estimate the SFR from the measured \Hb\
line flux, assuming a correction of $\mbox{I(\Ha)} = 2.86 \times \mbox{I(\Hb)}$.  This correction
factor represents the intrinsic line ratio for case B recombination at $10^4\K$ and an electron
density of $100\,\mbox{cm}^{-3}$ \citep{hummer-storey-87-caseB}. We use the relation $SFR(\Ha)
[\Msunyr] = 5.45\times10^{-42}L(\Ha) [\ergs]$ \citep{calzetti-etal-10-SFR-calibration} to convert
the measured luminosity $\mbox{L(\Ha)} = 6.07 \times 10^{40}\ergs$ into a SFR of
$\SFRnum\Msunyr$. Note that we have only included statistical errors in the SFR. Other systematic
errors in the calibration are likely to dominate the true error estimate. For example, we have no
ability to determine the internal reddening correction, which may increase the derived
star-formation rates by as much as a factor of 3, although the corrections are typically higher for
higher SFRs \citep{rosa-gonzalez-etal-02-SFR, calzetti-etal-07-midIR-SFR, hunter-etal-10-dwarf-SFR}.

We determined the oxygen metallicity of the star-forming regions in \galid\ using the R23
metallicity indicator \citep{pagel-etal-79-R23}, calibrated by
\citet{mcgaugh-91-R23-calibration}. R23 is defined as ([\OII] + [\OIII])/\Hb\ and is easily
measured. The R23 indicator has a well-known degeneracy and turnover at $\sim0.3\Zsun$ and large
systematic errors (on the order of 0.25\dex) due to age effects and stellar distributions
\citep{ercolano-etal-07-R23-problems}. Despite these drawbacks, it provides the only available
estimate of the galaxy metallicity from the spectrum of \galid. Unlike the SFR estimate, the R23
indicator is not as susceptible to internal reddening because the emission lines are close in
wavelength. Further, since \galid\ lies close to the turnover in the R23 indicator \citep[see, for
example,][Figure~12]{mcgaugh-91-R23-calibration}, even a significant amount of internal reddening
will have only a minor effect on the metallicity, and this will be much less than the intrinsic
systematic error.

We corrected the emission lines for foreground reddening using a standard Galactic reddening law
with $R_{v} = 3.1$ with $\EBV = 0.04$ \citep{schlegel-etal-98-dust}. Since our spectra do not show
[\NII] emission lines, we are unable to break the degeneracy in the R23 indicator, so we report both
upper and lower branch metallicities, obtaining the range $0.4 - 0.6\Zsun$. In calculating the
metallicity, we use the most recent solar Oxygen abundance of $12 + \mbox{log}(N_{O}/N_{H}) = 8.69$ from
\citet{asplund-etal-09-solar-abundance}. Table~\ref{tab: galaxy_properties} lists a summary of all
our measured and derived galaxy properties.

\begin{deluxetable}{p{1in}rrrcr}
\tabletypesize{\scriptsize}
\tablecaption{\label{tab: galaxy_properties}Galaxy Properties}
\tablewidth{0pt}
\tablehead{
  \colhead{} & 
  \colhead{}
}
\startdata
Galaxy ID  &  \galid \\
SDSS Name  &  J\,094330.670+053118.1 \\
\zgal      &  0.3529 $\pm$ 0.0001   \\
m$_{u}$    &  $22.5 \pm 0.5$ \\
m$_{g}$    &  $22.2 \pm 0.1$\\
m$_{r}$    &  $21.0 \pm 0.1$\\
m$_{i}$    &  $21.1 \pm 0.1$\\
m$_{z}$    &  $20.9 \pm 0.3$\\
M$_{r}$    &  $-20.4 \pm 0.1$\\
\Ls        &  0.4\,\Ls \\
\gmr       &  1.2   \\
\umr       &  1.5   \\
L(\Hb)     &  $2.12 \times 10^{40} \ergs$ \\
SFR (H$\alpha$) & $\SFRnum\Msunyr$   \\
12 + log(O/H)   & 8.48 (0.6\Zsun) Upper\\
12 + log(O/H)   & 8.26 (0.4\Zsun) Lower\\
\enddata
\tablecomments{Properties of galaxy at the approximate redshift of the absorption systems. The Galaxy
  ID is a polar co-ordinate system based on the position angle and radius (in arcsec), centered on
  the QSO position. Magnitudes are SDSS magnitudes corrected for MW dust extinction.}
\end{deluxetable}

\section{Discussion}
\label{sec: discussion}
\subsection{Galaxy-Absorber Connection}
\label{subsec: metal-poor}
{\bf Metal-poor cloud at $v = +365\kms$}\\
At a redshift of $\zgal = 0.3529$, the 19\arcsec\ angular distance between galaxy \galid\
corresponds to $\sim 95\kpc$. The velocity separation between the two components of the metal-poor
cloud and the galaxy is substantial ($\dv = +365, +445\kms$). By comparison, the Milky Way halo
contains clouds moving at radial velocities of up to $\sim450\kms$
\citep{wakker-vanwoerden-91-catalogue, putman-etal-02-catalogue, peek-etal-07-VHVC-distance}, so it
is not unreasonable that this cloud is associated with \galid. By analogy with the MW, we may
consider the main absorption low-metallicity infalling material. For this analogy to hold, we must
consider what both the observed SFR and the cloud metallicity can tell us. Let us consider these two
issues in turn.

If we consider a scenario in which this cloud is part of an outflow from the main galaxy, we must
reconcile both the low SFR in the host galaxy \galid, and consider the metallicities. The
travel time to an assumed distance $d > 95\kpc$ at the observed velocity difference $365\kms$ is
some $\sim250Myr$. Star-formation induced outflows can be driven by the correlated supernovae of
early type stars in OB associations \citep{mac-low-etal-88-superbubbles}. Since those stars have
typical lifetimes of only $\simlt 10\Myr$ \citep{meynet-etal-94-SFR-grids}, much shorter than the
minimum travel time to the observed distance of 95\kpc, the current observed SFR
does not provide a constraint on the gas origin.

Next let us consider the metallicity. The observed \HI\ and \CIII\ column densities in both strong
components are the same (to within $0.1-0.2\dex$), with $\logNHI \sim 16$ and $\logNCIII \sim
13.5$. At these \HI\ column densities, we should expect to see appreciable absorption columns from
metal lines in gas enriched to greater than a few tenths solar. Indeed, we see similar strength of
absorption in \CIII\ in the \OVI\ cloud as the metal-poor cloud, but the \OVI\ cloud has more than
$25\times$ lower \NHI. This points to the possibility that the cloud is metal-poor, but is not
conclusive, since ionization corrections may be significant.

To quantify this assertion of low metallicity, we attempted simple ionization modeling. The relative
lack of metal lines makes detailed modeling impossible, but the \CIII\ detection, as well as the
\CII, \MgII\ and \OVI\ upper limits gives us some direction. We examined grids of {\it Cloudy}
photoionization models \citep[version 08.00;\ ][]{ferland-etal-98-cloudy}, varying both the
metallicity and the dimensionless ionization parameter $\logU$ (the ratio of the density of hydrogen
ionizing photons to total hydrogen density), integrating to a fixed $\logNHI = 16.0$. We used the
updated ``HM05'' UV background spectrum, which is the average UV background produced by AGN and
starburst galaxies \citep{haardt-madau-96-UV-background}; we did not include a contribution from
nearby star-formation. \citet{tumlinson-etal-11-J1009-LLS} give a relation to derive the
contribution of star-forming regions to the UV background, as a function of escape fraction,
distance and SFR. At a distance of $d > 95\kpc$ and $\mbox{SFR} = \SFRnum\,\Myr$, star formation
contributes to the ionization field at the level of a few percent for reasonable values of the
escape fraction of ionizing photons \citep[i.e. $f_{esc} < 0.04$;][]{putman-etal-03-Halpha,
  chen-etal-07-GRB-fesc}. Note also that photoionization of \OV\ to \OVI\ requires photons with
energies of 114\eV\ or greater, which cannot be supplied by star-formation.

\begin{figure}
  \epsscale{1.2}
  \plotone{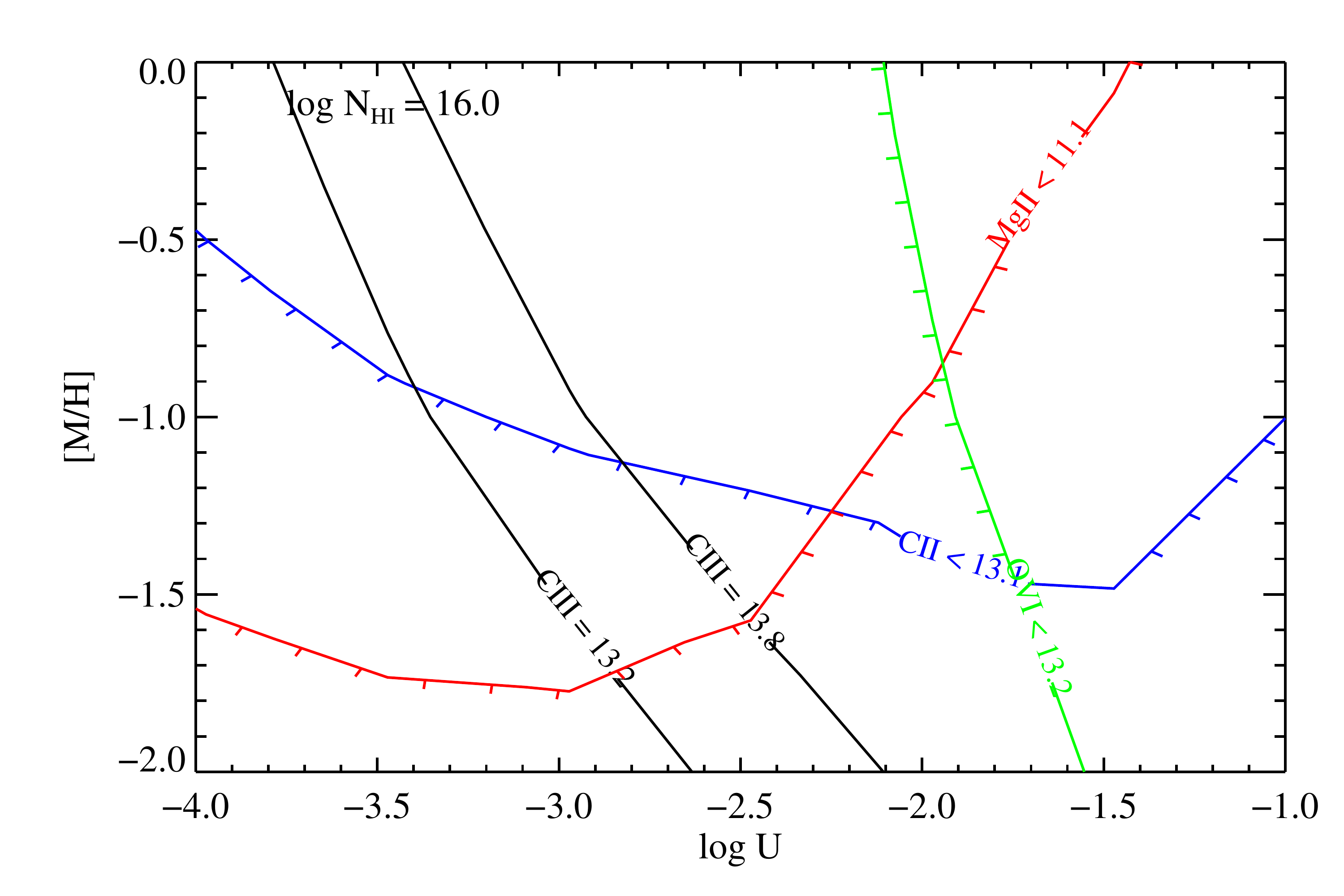}
  \caption{Column contours predictions from photoionization model of the metal-poor cloud. Limits
    are given based on the measurements from Table~\ref{tab: column_densities}. Small tick marks on
    the contours indicate the allowed region of parameter space for upper limits.}
  \label{fig: metal_poor}
\end{figure}

Figure~\ref{fig: metal_poor} shows contours of the predicted column densities in metallicity-\logU\
space for these cloudy models. As can be seen, the \MgII\ and \CII\ non-detections provide tight
constraints on the metallicity, although this restriction relaxes at high ionization parameter (low
density). The \OVI\ non-detection limits the allowed region of parameter space to $\logU < -1.5$,
but the combined detection of \CIII\ and non-detection of \MgII\ is the most restrictive. These
combine to limit the metallicity of the cloud to $\MH < -1.5$. At this point, it is worth
emphasizing the value of the follow-up HIRES optical spectrum. Without it, the metallicity would be
significantly less constrained, and our interpretation much less certain.

The combination of the high velocity relative to the galaxy, the low metallicity, and the large
distance from the galaxy leads us to the conclusion that the halo cloud at $v = +365-445\kms$ is
an accreting, metal-poor gas cloud, providing fuel for future star formation in galaxy \galid.  
\newline

{\bf \OVI\ cloud at $v = 85\kms$}\\
\label{subsec: OVI-cloud}
The velocity separation between the \OVI\ cloud and galaxy is $\dv = 85\kms$, which is typically
close enough to consider the gas to exist in the galaxy halo. It is interesting to note that a cloud
with such a low velocity in the MW halo may not even be classified as a halo cloud, given its low
velocity separation from the galaxy. Rather, the \OVI\ cloud would likely be considered part of the
$\sim 3\kpc$ high, ionized gas surrounding the thick disk
\citep[c.f.][]{savage-etal-03-OVI-disk-halo}, as opposed to a distant halo cloud. The detection of
\OVI\ at a distance $d \geq 95\kpc$ from the disk may have implications for the interpretation of
the MW \OVI\ halo clouds.

To interpret the \OVI\ cloud, we considered the feasibility of several scenarios to explain physical
conditions of the cloud. In doing so, we adopt the single component metal-line column densities from
the Voigt profile fitting in the previous section (Table~\ref{tab: profile_fitting}). The principle
source of error, which propagates to our derived physical quantities, is the large permitted range
of \CIII\ column density.

A pure collisional ionization equilibrium (CIE) model can be quickly excluded by examination of the
absorption line widths and column density. \OVI\ traces collisionally ionized gas in the range
$\logT = 5.2 - 6.5$, peaking at $\logT = 5.5$ \citep{sutherland-dopita-93-gas-cooling,
  gnat-sternberg-07-non-equilibrium-cooling}. \OVI\ and \CIII\ may coexist at the low end of this
range, but the \NOVI\ to \NCIII\ ratio is a very steep function of temperature; the measured range
in the column density ratio $\NOVI/\NCIII = 0.6 - 1.7$ allows temperatures only in the narrow range
$\logT = 5.25 - 5.3$.  At this temperature, the implied thermal doppler line-width for \HI\ is
$b_{therm} = \sqrt{2kT/m} = 57\kms$, or a full-width at half-maximum width of 94\kms. This line
width is clearly excluded by a visual inspection of the data; a Voigt profile fit (accounting for
the COS line-spread function) yields $b_{\scHI} \simeq 32\kms$, confirming that a pure collisional
ionization equilibrium model is ruled out.

The detection of both \CIII\ and \OVI, and the limits on \CII\ and \MgII\ column densities give us
some handle on a simple photoionization scenario. Simple plane-parallel {\it Cloudy} photoionization
models with the canonical HM05 spectrum can reproduce the observed metal line column densities and
line ratios. The ratio of \NOVI\ to \NCIII\ fixes the range of allowed ionization parameter (and
hence density), since the column density ratio is independent of metallicity for a fixed abundance
pattern, and the UV background is also fixed. The allowed metallicity range can then be obtained
using the permitted density ranges, and varying the metallicity until the observed column densities
are obtained. 

\begin{figure}
  \epsscale{1.2}
  \plotone{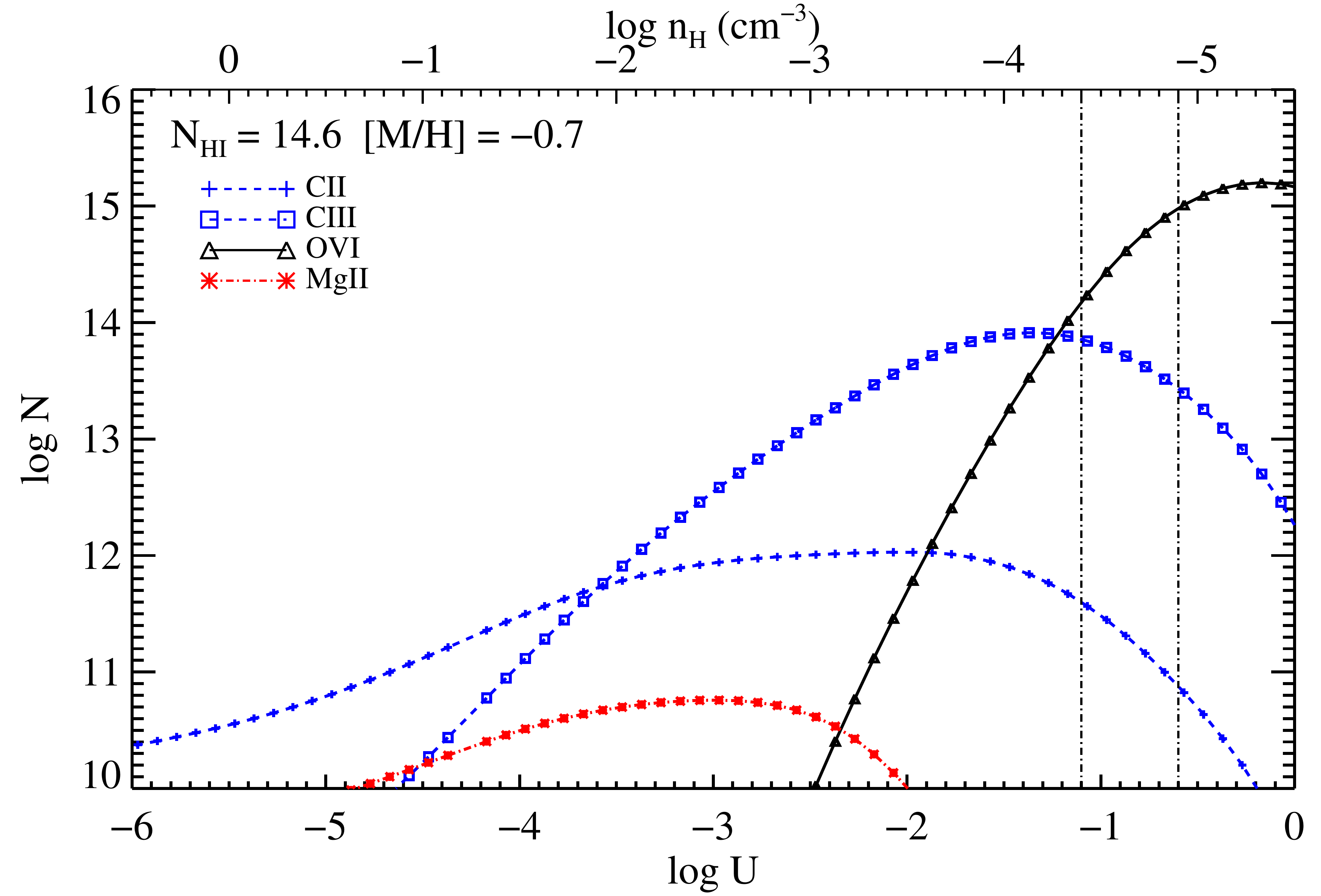}
  \caption{Column density predictions from photoionization model of the \OVI\ cloud. The model
    assumed a plane-parallel geometry, illuminated by the HM05 UV background spectrum, integrated to
    $\NHI = 10^{14.6}$ with $\MH = -0.7$. Vertical dashed lines show the region allowed by the
    observed \OVI\ to \CIII\ column density ratio.}
  \label{fig: photo-plot}
\end{figure}

The observed conditions are reproduced for a model with $\logU = -1.1 - -0.4$ and metallicity in the
range $\MH =-0.3 - -1.1$ for a solar abundance pattern. Given the input HM05 spectrum is fixed at
all points (and hence the number of ionizing photons is also fixed), this $\logU$ range corresponds
to a density range $\loghden = -4.4 - -4.9$. However, this assumes both the shape and normalization
of the UV spectrum is known and equal to the HM05 spectrum (which is simply the {\it average}
spectrum); the true normalization and shape are still quite uncertain
\citep[c.f.][]{scott-etal-02-UV-background, shull-etal-99-UV-background}. We have no ability to
determine the true shape or normalization of the radiation field (i.e.\ our data do not show any
density diagnostics), so we can only proceed using the canonical UV background field; nevertheless,
this should be kept in mind. A plot of the predicted metal line column densities as a function of
\logU\ and \hden\ is shown in Figure~\ref{fig: photo-plot}. To construct this plot, we took the
metallicity to be $\MH = -0.7$, in the middle of the allowed range. The vertical dashed lines mark
the range of density/ionization parameter permitted by the \OVI\ to \CIII\ column density ratio.

At these densities, the fraction of hydrogen in neutral form is extremely low ($10^{-4.5 -
  -5.1}$). To obtain the observed neutral hydrogen column density of $\logNHI = 14.6$ at this
ionization fraction and physical density would require a cloud with a length scale $L \sim
0.1-1.2\Mpc$.  A cloud with this physical size is unlikely to be a distinct cloud in a galaxy halo,
but the length scale is consistent with absorption arising in a large-scale filamentary structure
(i.e. the \Lya\ forest) or a galaxy group \citep[see, for
example][]{schaye-01-Lya-forest}. Intragroup gas has separately been suggested as a possible origin
for highly ionized species such as \OVI\ \citep{mulchaey-etal-96-hot-gas} and a few examples have
been observed, both in the Local Group and the low-$z$ IGM
\citep{tripp-savage-00-PG0953+415_z0.1423, nicastro-etal-03-nature}.

In order to assess a possible group origin, we examined galaxies in the field near \galid\ using
available SDSS data. We took the photometric redshift estimates\footnote{Photometric redshifts are
  especially challenging to accurately determine at $z = 0.35$, due to the 4000$\A$ break moving
  from the g to r bands, and the $p(z)$ distributions of these systems are correspondingly broad.}
for galaxies in the surrounding region as given by \citet{cunha-etal-09-SDSS-photoz}. The closest
potential group candidate is a set of three galaxies $\sim 1\Mpc$ to the NNW: all are brighter than
$0.4\,\Ls$ if they are at the redshift of the absorber, and have photometric redshift probability
distributions $p(z)$ broadly consistent with $z=0.35$. The $0.4\,\Ls$ level is typically used in
structure-finding algorithms, exploiting the red sequence of galaxies in large virialized structures
such as groups and clusters \citep[see, e.g.][]{koester-etal-07-maxBCG}; the fact that \galid\ is
also 0.4\Ls\ is coincidental.

With such a small number of galaxies, the radius of such a system would be only $r_{200} \sim
0.4\Mpc$ \citep{hansen-etal-09-clusters-groups} and would not encompass \galid. There are a number
of galaxies in the vicinity of \galid\ which have photometric redshifts broadly consistent with
$z=0.35$, but all are fainter than $0.4\,\Ls$. These galaxies lie primarily to the S.  We thus
conclude that \galid\ probably does not exist in a galaxy group, and that the \OVI\ does not
therefore arise from the group environment. This conclusion is also consistent with the above
finding that CIE models do no adequately describe the data. We are thus left with the possibility
that the \OVI\ arises in a filament. Spectroscopic confirmation of these faint galaxies is required
to further investigate the filamentary structure hypothesis.

To check the physical consistency of the filament scenario, we compared the measured properties of
the \OVI\ cloud to the IGM equation of state predictions of
\citet{ricotti-etal-00-IGM-eqn-state}. At a density $\hden = -4.4 - -4.9$, the cosmic overdensity is
$\delta \simeq 12-37$. Using this overdensity and the effective IGM equation of state (EOS), $T = T_0
(1+\delta)^{\gamma-1}$ from \citet{ricotti-etal-00-IGM-eqn-state}, we can derive the implied thermal
Doppler parameter for \HI. We interpolate between redshifts in their Table 6 to obtain EOS
parameters $T_0 = 7.1 \times 10^3$ K and $\gamma - 1 = 0.75$ for $z = 0.4$. This resulting EOS then
predicts temperatures in the range $\logT = 4.6-5.0\K$ and $b_{t} = (2kT/m)^{0.5} = 28-43\kms$.

\begin{figure}
  \epsscale{1.2}
  \plotone{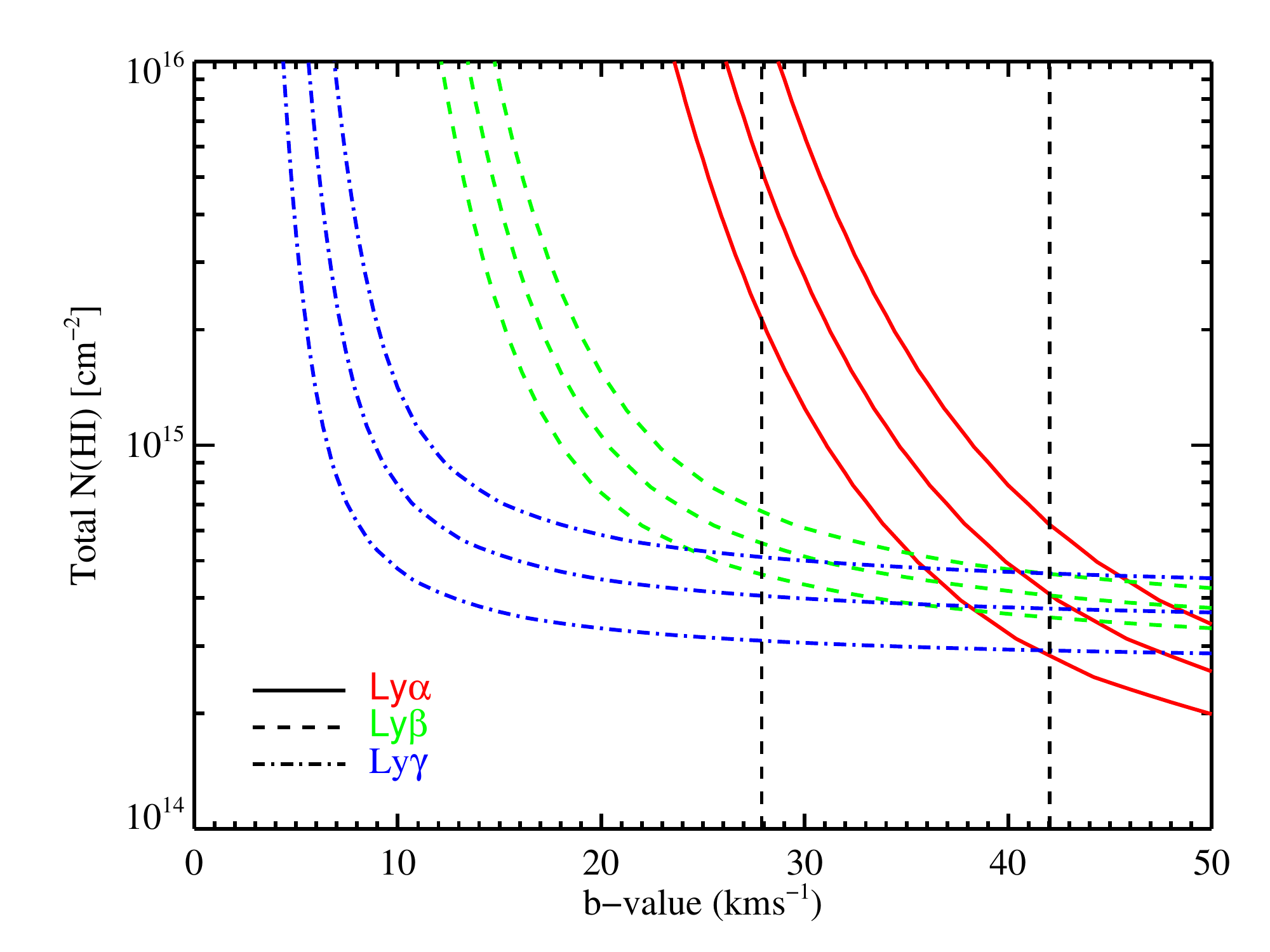}
  \caption{Curve of growth for the \HI\ lines in the \OVI\ cloud. The curves show the measured value
    and $\pm 1\sigma$ error bars. The vertical dashed line shows the prediction from the IGM
    equation of state. See text for details.}
  \label{fig: cog}
\end{figure}

To compare to this predicted Doppler parameter with the measurements, we performed a curve of growth
analysis on the 3 strongest Lyman series lines in the \OVI\ cloud. The results are shown in
Figure~\ref{fig: cog}. The vertical dashed lines show the Doppler parameter predicted by the IGM
EOS. The curve of growth shows a broad region of allowed Doppler parameters, which are consistent
with the theory predictions.  We conclude that the absorber has a density, ionization state, cloud
size, and line broadening consistent with a filament of the large scale structure. While we observe
that some galaxies in the region around \galid\ have photometric redshift estimates consistent with
$z=0.35$, a more detailed spectroscopic survey of the large-scale galaxy distribution around \galid\
is necessary to confirm this suggestion. In addition, the principle source of error in constraining
the physical conditions of the \OVI\ cloud is the large range of column densities allowed by the
fits to the \CIII\ line. A higher resolution and/or higher signal-to-noise spectrum would allow us
to be more precise about the physical conditions, but is unlikely to alter the broad conclusions.

\section{Conclusion}
We have presented the observations of two gaseous structures near or in the halo of a galaxy at $z =
0.35$.  Absorption lines due to an enriched gas filament and infalling metal-poor cloud are detected
with small velocity offsets from the systemic galaxy redshift. The metal-poor cloud has strong \HI\
absorption ($\sim 10^{16}\cmsq$) and weak metal lines, with only weak lines of \CIII. A sensitive
non-detection of \MgII\ in the optical high resolution spectrum, combined with the observed \CIII\
line strengths, indicates this cloud is metal-poor ($\MH < -1.5$).

The other absorber contains strong \OVI\ absorption, and is consistent with arising in cool
photoionized gas at modest overdensities ($\loghden = -4.4 - -4.9; \delta \simeq 12-37$). At these
densities, the implied length scale for the \HI\ is $\sim 0.1-1.2\Mpc$ leading to the suggestion
that the gas arises in a group or filament. At an impact parameter $\rho = 95\kpc$ and velocity
separation $\dv = 85\kms$, a casual inspection would reasonably associate the \OVI\ absorption with
the galaxy halo; this example serves as a cautionary tale against such association. Inspection of
photometric redshifts for nearby galaxies reveals several faint galaxies consistent with being at $z
= 0.35$, but there are no bright galaxies ($L > 0.4\,\Ls$ at $z=0.35$) nearby, which argues against
the gas arising in an intragroup environment, and hence the gas most likely exists in a
filament. Spectroscopic followup of galaxies in the field is required to confirm this conclusion.

In our analysis of both clouds, we detected a mild level of unresolved saturation. This is likely to
occur in many low signal-to-noise data ($S/N < 10$), and is likely to be especially problematic for
metal lines, which are intrinsically narrow.

When complete, our survey will provide some 80 QSO sightlines similar to those presented here, that
are selected to pass through the halos of galaxies at a range of luminosities, colors and impact
parameters. Observations such as these are required to unravel the connection between gas and
galaxies, an increasingly important part of the galaxy evolution puzzle. They may also provide
context for the interpretation of observations of the Milky Way halo.

\acknowledgements

C.~T. thanks Sarah Hansen for helpful discussions on the selection of galaxy groups from photometric
redshift data. We thank Derck Massa for providing early copies of the COS flat field corrections,
and Edward Jenkins for his careful processing of the flats. Support for program GO\,11598 was
provided by NASA through a grant from the Space Telescope Science Institute, which is operated by
the Association of Universities for Research in Astronomy, Inc., under NASA contract NAS
5-26555. J.~X.~P. and J.~W. acknowledge partial support from NSF CAREER grant AST--0548180. Some of
the data presented herein were obtained at the W.M. Keck Observatory, which is operated as a
scientific partnership among the California Institute of Technology, the University of California
and the National Aeronautics and Space Administration. The Observatory was made possible by the
generous financial support of the W.M. Keck Foundation. The authors wish to recognize and
acknowledge the very significant cultural role and reverence that the summit of Mauna Kea has always
had within the indigenous Hawaiian community.  We are most fortunate to have the opportunity to
conduct observations from this mountain.

\facility{HST(COS); Keck(LRIS); Keck(HIRES)}

\bibliography{MASTER}
\bibliographystyle{apj}

\clearpage


\end{document}